# Laser-guided acoustic tweezers

Qing Wang, Shuhan Chen, Jia Zhou*, Antoine Riaud*

Acoustic tweezers[1–3] can manipulate microscopic objects and cells independently of the optical[4,5], magnetic[6,7] and electrical properties [8,9] of the objects or their medium. However, because ultrasonic waves are attenuated within few millimeters, existing devices must synthesize intricate and powerful acoustic fields in a very narrow footprint immediately close to the manipulated object [10–14]. Here we show that the design of microscale acoustic tweezers can be considerably simplified by taking advantage of the nonlinear nature of the acoustic trapping force. In our experiment, a featureless piezoelectric crystal coated with a photoacoustic conversion layer is hit by an electric pulse and a spatially modulated laser pulse to generate synchronized electro- and photo- acoustic waves. Interference between these waves creates a hybrid acoustic trapping force 30 times stronger than the laser pulse alone but with the same spatial information.  By disentangling the tradeoff between spatial resolution and trapping force that has long held back the development of acoustic tweezers, this field hybridization strategy opens new avenues for cell manipulation in organ on chip and organ printing.

## Main

Techniques for precisely manipulating microscopic objects are invaluable in life science from fundamental biology [15] to medical applications [16]. Combined with holographic techniques [5], optical tweezers can manipulate multiple particles independently with a high dexterity. However, capturing particles using optical radiation force (as with optical tweezers) requires power densities harmful to cells, which has limited the use of optical tweezers mainly to subcellular components. An alternative to optical tweezers is acoustic tweezers which can apply $10^5$ larger forces than optical tweezers at equal power density.

Acoustic tweezers capture particles by using the radiation pressure of sound. For instance, **Figure 1**.a. shows the generation of a plane acoustic wave (thereafter referred as Z-wave) by a typical acoustic tweezers placed under the particle. Under the action of the incident wave, the particle is forced to vibrate (**Figure 1**.b). For particles much smaller than the acoustic wavelength, these vibrations mainly consist of cycles of compression and expansion (monopolar) and back-and-forth oscillations (dipolar). In this paper, it is helpful to adopt a quantum perspective where acoustic excitation of a small particle creates a pseudo-particle akin to an atom (where the particle would be the nucleus and the vibrations would be the electron) that we will call vibron. Similar to atoms, parity-time symmetry [17] (Newton third law) shows that such vibrations do not generate a steady motion or steady force by themselves. However, in addition to making the particle vibrate, the incident acoustic field also breaks the symmetry of these oscillations, resulting in the acoustic radiation force experienced by the particle [18]. In the quantum analogy, a vibron exposed to an acoustic field is endowed with an acoustic radiation potential (Gor'kov potential) that depends on the overlap between the vibron and the incident field. Then the radiation force is the gradient of this radiation potential.

Most of the current acoustic tweezers rely on a piezoelectric transducer (as shown in **Figure 1**.a) to generate the incident acoustic wave. The nearly lossless conversion of electrical to acoustic energy allows generating strong acoustic fields with minimal heating. However, such devices are limited by a trade-off between resolution and refreshing rate: fast transducer microarrays are complex and can only combine a few hundred elements at best [1,3,12], microbubbles arrays (that can be generated by laser [14,19] or hydrolysis [11,20]) have a limited resolution and a refreshing rate of 0.1 fps and high-resolution holographic plates [2] or electrodes [10,21] with up to $10^4$ pixels are not reconfigurable at all.

An alternative to piezoelectric actuation is using laser pulses to generate acoustic waves (**Figure 1**.c). In this photoacoustic conversion, a laser pulse causes transient heating of a light-absorbing material, which induces thermal expansion to generate an acoustic wave (thereafter referred as L-wave). The principal advantage of using light for ultrasound generation is that optical images formed at macroscale (for instance by masks [22], or spatial light modulators [23,24]) are readily shrunk to microscale using a microscope objective (see Methods 2.3). This allows forming arbitrary acoustic fields at microscale from optical images with a 100 times better than acoustic holograms, and a 100 times faster refreshing rate than acoustic devices based on bubbles. Yet, despite recent progress in material synthesis, photoacoustic conversion efficiency remains as low as $10^{-7}$ for a $10^4$ W/cm$^2$ laser fluence (Method), so that the vibrations of the particle are much fainter (**Figure 1**.c) than those obtained from lossless electroacoustic conversion, which yields an acoustic radiation force even weaker than the optical radiation force from the same laser pulse (**Figure 1**.f)[25].

Here, we show that the nonlinear P-T symmetry breaking at the heart of the acoustic radiation force allows to considerably simplify the design of acoustic tweezers. This is illustrated in **Figure 1**. In existing acoustic tweezers, the wave that drives the vibrations of the particle is the same as the one to break their symmetry. This needs not to be the case. We consider here a superposition of a strong electroacoustic plane wave (Z-field) to drive the oscillations of the particle, and a weaker photoacoustic L-field to break their symmetry (**Figure 1**.e). In the quantum analogy, this is akin to forming a molecular bond by hybridization of atomic orbitals between the L and Z fields. The resulting force retains the high spatial resolution and frame rate of the light pattern used in photoacoustic generation, but is amplified more than 30 times by the electroacoustic field. For the first time, this gain in efficiency allows the radiation pressure of photoacoustic devices to exceed that of optical tweezers at equal power density (**Figure 1**.f) (See Methods 3 for the characterization of the manipulation force for each type of device). Furthermore, by analogy with bonding and antibonding molecular orbitals, adjusting the relative phase between the Z and L fields inverts the symmetry-breaking effect of the L-field, which reverses the direction of the force (switching from a pulling to a pushing force), a phenomenon that has no analog in conventional acoustic tweezers.

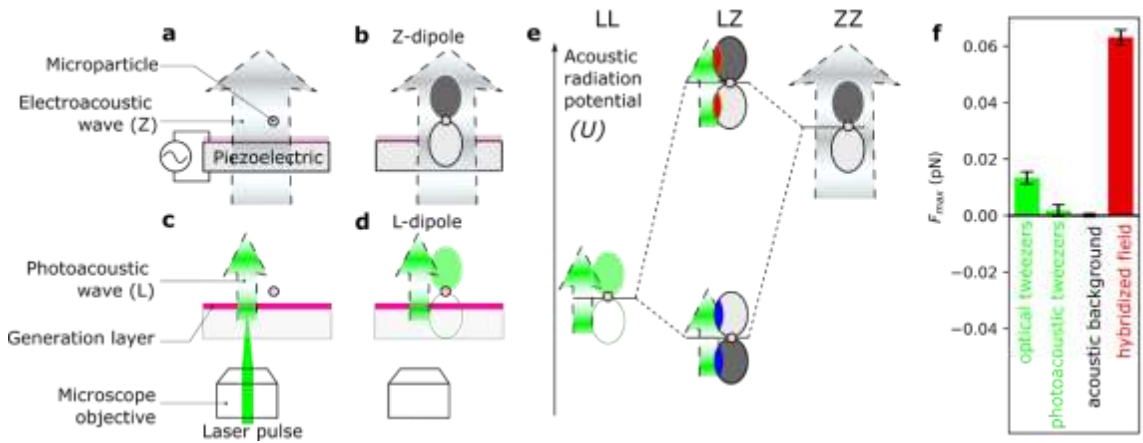

**Figure 1: Amplification of the radiation force of a photoacoustic wave by hybridization with an electroacoustic wave.** (a) generation of a spatially-uniform acoustic field (Z) by piezoelectric conversion, and resulting the creation of a vibron with monopolar and dipolar oscillations, represented by the two lobes (b). The acoustic field (arrow) is uniform, so the overlap between incident field and particle vibration is constant, and therefore the vibrating particle experiences no preferential motion in the (x,y) plane. (c) Generation of a localized acoustic wave (L) by photoacoustic conversion, and resulting the creation of a vibron with monopolar and dipolar oscillations (d). Due to poor photoacoustic conversion efficiency, the photoacoustic field is faint and the particle does not vibrate much, resulting in a small overlap and low force. However, the L-field is relatively focused (narrow arrow), which allows to precisely decide the location of the force. (e) Hybridization of electroacoustic and photoacoustic fields. The particle oscillations induced by the Z-field are coupled to the symmetry breaking of the L-field. In the quantum analogy, the Z-vibron is hybridized with the L-field, which results in much stronger radiation potential (LZ). Similar to bonding and antibonding molecular orbitals, the hybridized acoustic radiation force can be reversed depending on the relative phase difference between the Z and L fields. (f) experimental measurements of several forces acting on the particle using the same laser and same power density.

A quantitative picture can be gained by using the radiation force theory for small spheres, which was recently extended to pulses such as those generated by photoacoustics conversion [26]. The radiation force reads $F = -V_p \nabla U$, with $U = \frac{1}{2} f_1 \frac{\langle p^2 \rangle}{\rho_0 c_0^2} - \frac{3}{4} f_2 \rho_0 \langle v^2 \rangle$ the acoustic radiation potential (Gor'kov potential [27]). Here, $V_p$ is the particle volume, $\rho_0$ and $c_0$ are the fluid density and sound velocity, $p$ and $v$ are the total incident acoustic pressure and vibration velocity. If two acoustic fields labeled (L) and (Z) interfere, the total incident pressure reads $p = p^{(L)} + p^{(Z)}$ and the total incident velocity $v = v^{(L)} + v^{(Z)}$. The potential now includes crossed-terms and reads $U = U^{(ZZ)} + U^{(ZL)} + U^{(LL)}$. If the Z-wave is a plane traveling wave $p^{(Z)} = A^{(Z)} \cos(k_z z - \omega\tau - \omega t)$, with $\tau$ an adjustable delay between the L and Z fields. $U^{(ZZ)}$ is invariant in space and therefore exerts no force due to its vanishing gradient. Furthermore, due to the small photoacoustic conversion efficiency, $U^{(LL)}$ is negligible. Upon derivation of the potential (see Methods 1 for the detailed process), the x-component of the force reads (a similar expression is obtained for y):

$$F_x = -V_p A^{(Z)} \cos(\omega\tau - \phi_x) f_x^{(L)} \qquad 1$$

The function $f_x^{(L)}(x, y, z)$ and the phase $\phi_x(x, y, z)$ carry the spatial information provided by the light-pattern (L-field). The amplitude of the force is proportional to $A^{(Z)}$, which acts here as a gain on the radiation force from the L-field. Finally, the direction of the force can be reversed by adjusting the delay $\tau$.

The experimental setup, shown in **Figure 2**.a, is built around an inverted fluorescence microscope. The particles to be manipulated are suspended in a saline solution to minimize sedimentation effects. We note that a range of method based on electrophoresis would fail to achieve manipulation is such a conductive solution or cell media [8,28]. Similar to other acoustic tweezers, the dispersion is injected into a microfluidic channel in order to minimize disruption by acoustic reflections and external air currents. The tweezers themselves are a featureless composite (**Figure 2**.b) consisting of a transparent piezoelectric transducer [29,30] and a photoacoustic conversion layer with a strong absorption peak in the 532 nm region [31]. Therefore, the composite is transparent to red and blue and can generate acoustic waves by absorbing light pulses from a green laser. Before reaching the microscope, the laser beam wavefront is adjusted by a holographic setup similar to those used in optical tweezers. The light pattern is visualized using fluorescent nanoparticles dispersed in the fluid. The photoacoustic pulse generated by the laser was measured in-situ and yields a displacement at a quasi-constant speed of 4.9 cm/s. The piezoelectric is excited by short wave trains (5 periods) to generate 1.7 nm amplitude z-propagating spatially-uniform plane wave. Respective acoustic waveforms and scanned fields are shown in Methods 2.3 and 2.4.

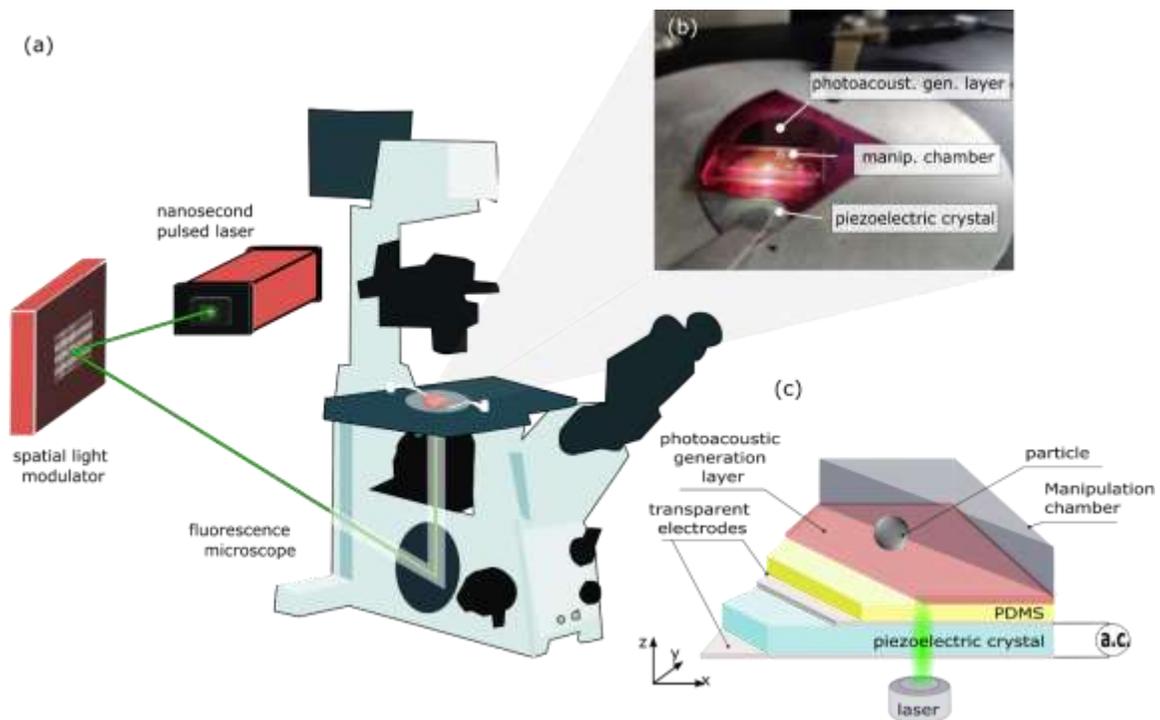

**Figure 2: Experimental setup.** (a) generation of a microscale optical light pattern using a spatial light modulator and a fluorescence microscope (b). Experimental image of the laser-guided acoustic tweezers. (c) Cross section of the laser-guided acoustic tweezers. A piezoelectric crystal (LiNbO$_3$) sandwiched by transparent electrodes generates the electroacoustic wave. A composite material (PDMS, generation layer) converts light pulses (laser) into acoustic waves by photoacoustic conversion. The interference of those waves enables the manipulation of particles contained in the manipulation chamber. The manipulation chamber is a square of 1 mm edge length and 200 μm depth.

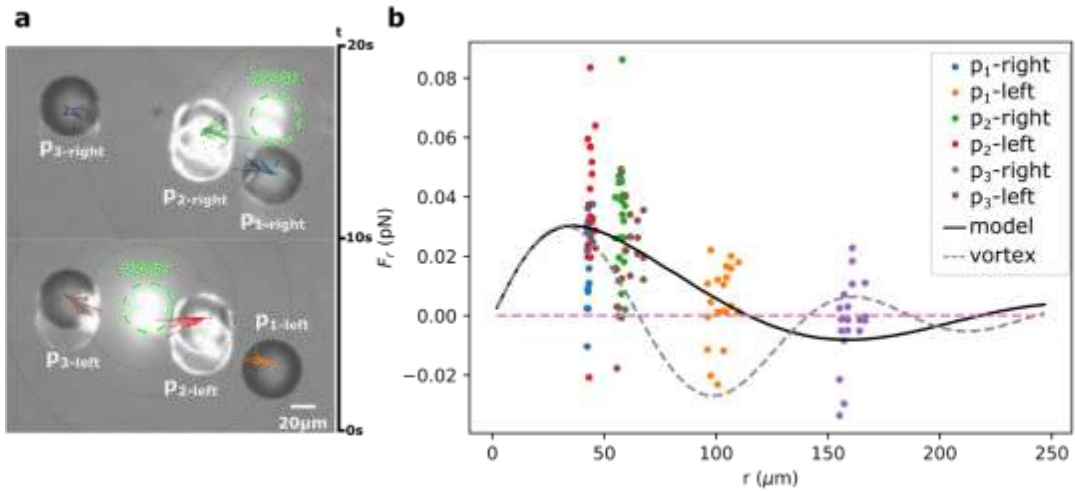

**Figure 3: Dynamic manipulation of particles and spatial distribution of the trapping force.** (a) Using the spatial light modulator, the laser beam is periodically moved from left to right. Particle motion over several cycles is indicated by the arrows. The image is doubled due to the birefringence of the LiNbO$_3$ crystal. (b). Estimated trapping force depending on the distance of the particles from the laser spot. The model (solid line) is obtained by finite element modeling and fitted to the experimental datapoints. "vortex" refers to the theoretical limit of state-of-the-art selective acoustic tweezers using spherical acoustic vortices generated by holography.

In order to get an estimate of radiation force on the particles, we study the displacement speed of several particles under the action of the laser-guided tweezers depending on the relative distance between the particles and the laser spot **Figure 3**.a. This displacement is opposed by hydrodynamic and friction effects. In this regard, the Stokes drag is a lower-bound estimate of the force acting on the particles, and can be estimated from the displacement speed of the particles (see Methods 3 and 4 for a more extensive description of the measurement and underlying assumption). The force profile is obtained by recording the motion of several particles within a 150 µm radius of the laser spot. Our datapoints shown in **Figure 3**.b are fitted with the force profile by a finite element model of the acoustic radiation force theory for pulses [26] (see Methods 1 and 5 for the main equations). The results, shown with the solid line, are in qualitative agreement with the experiment. The force on the particles depends on the distance to the laser spot, with a peak for particles located at 40 µm from the laser spot. The 40 µm radius for the maximum trapping force is exactly the same as the theoretical limit state-of-the-art holographic acoustic tweezers [10], but secondary positive and negative overshoots are suppressed. Consequently, particles located 100 µm from the primary acoustic trap would be repelled by holographic tweezers but are not affected by the laser-guided tweezers, which allows a more accurate manipulation. The maximum force at this peak (0.06 pN) is nearly 5 times that of optical tweezers and more than 30 times that of photoacoustic tweezers at equal power density. However, the experimentally measured force is only 4.5% of that predicted by the model. This discrepancy will be discussed later in the paper.

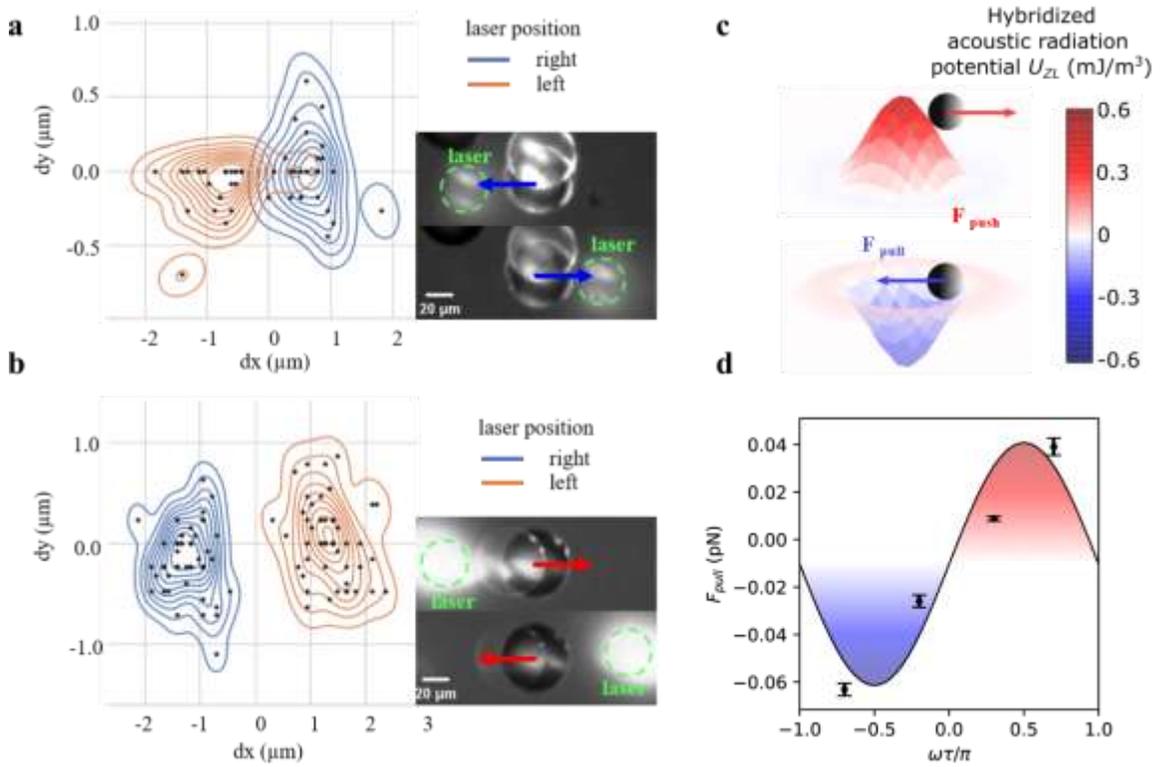

**Figure 4: Reversal of the trapping force by adjusting the relative phase between electroacoustic (Z) and photoacoustic (L) fields.** The phase difference is composed of a constant delay (1.3 π) set by the experimental hardware and an adjustable phase difference $\omega\tau$. Only the adjustable phase $\omega\tau$ is shown in the figures. Using the spatial light modulator, the laser beam is periodically moved from left to right. The particle motion is tracked over several cycles. (a,b) particle displacement in (x,y) coordinates (the contour plots are a guide for the eye) for (a) $\omega\tau = -\pi/2$ and (b) $\omega\tau = \pi/2$. The tweezers commute between pulling and pushing mode. (c) Simulated acoustic radiation potential for $\omega\tau = -\pi/2$ (pull) and $\omega\tau = \pi/2$ (push). We note that $U_{ZL}$ is considerably smaller than $U_{ZZ} \approx 22$ mJ/m³ (d) Variation of the radiation force depending on $\omega\tau$. Experimental data are shown as error bar brackets, while fitted simulation results are shown as the solid line. The slight negative offset is due to the LL component of the force. The colors are guide for the eye to indicate the sign of the ZL component. In microscopy photographs (inset of a,b), the image is doubled due to the birefringence of the LiNbO₃ crystal.

Another key feature predicted by Eq. 1 is that the amplified radiation force can be reversed if the acoustic pressure of the Z-field has an opposite sign with the L-field (**Figure 1**.e, **Figure 4**.c). This is achieved by introducing a delay between the triggers of the laser and the piezoelectric in order to adjust the relative phase difference between the two fields. **Figure 4**.a,b shows the displacement of the particle for two different delays. In **Figure 4**.a, the particle is pulled toward the laser beam, whereas in **Figure 4**.b it is pushed away from it. This trend is confirmed over a range of delays by experiments and numerical simulations (**Figure 4**.d) with the sinusoidal trend predicted in Eq. 1. This is a distinctive feature from conventional acoustic tweezers wherein the pulling or pushing nature of the radiation force was solely controlled by material properties, so one had to resort to using complex helical acoustic fields [10].

Although model and experiments agree qualitatively on the spatial profile of the force and the reversal of the force depending on the phase difference between the photoacoustic and electroacoustic waves, the experimental force estimated from the Stokes drag is only 4.5% of that predicted by the model. The Stokes drag is just one of several forces opposing the particle motion (see Methods 3), therefore this discrepancy is most likely due to other forces such as the effect of the channel walls, that can increase the drag coefficient by up to 8 times [32], and to adhesion forces between the PDMS and the spheres that must be overcome before the sphere is set in motion. A more accurate measurement of the hybridized radiation force, for instance using optical tweezers, could help pinpoint the origin of the discrepancy.

Even though the trapping power of this first implementation is low, photoacoustic conversion efficiency can be increased by several orders of magnitude by using stronger pulses with a lower pulse repetition frequency. On the electroacoustic front, the voltage applied to the LiNbO$_3$ crystal can be increased by two orders of magnitude before reaching dielectric breakdown. Such high voltage may be supplied by specialized pulsed amplifiers. It is therefore likely that with suitable improvements, laser-guided acoustic tweezers can deliver forces in the nN or higher. Finally, while this method was demonstrated in the case of acoustic waves, the use of external fields to amplify nonlinear forces seems readily applicable to other fields, such as electrophoresis (where fluid acceleration would be the electric field and the pressure the electric potential) and electromagnetism (where the Minkowski stress tensor is analog to the Brillouin stress tensor)[33].

# Methods

### 1. Calculation of the acoustic radiation force

In our previous study [26], we have shown that the Gor'kov equation is relevant to calculate the radiation force of transient acoustic fields on small spheres and reads:

$$F = -V_p \nabla U, \qquad 2$$

with the acoustic radiation potential:

$$U = \frac{1}{2} f_1 \frac{\langle p^2 \rangle}{\rho_0 c_0^2} - \frac{3}{4} f_2 \rho_0 \langle v^2 \rangle \qquad 3$$

Here, $p$ and $v$ are the total acoustic pressure and vibration velocity. If two acoustic fields labeled (L) and (Z) interfere, the total pressure reads $p = p^{(L)} + p^{(Z)}$ and the total velocity $v = v^{(L)} + v^{(Z)}$. The potential now reads $U = U^{(ZZ)} + U^{(ZL)} + U^{(LL)}$, where the superscripts indicates the origin of the cross-terms. Therefore, $U^{(ZL)}$ mixes the L-field (light-pattern information) and the Z-field (piezoelectric power).

If the Z-wave is a plane traveling wave $p^{(Z)} = A^{(Z)} \cos(k_z z + \omega\tau - \omega t)$, with $\tau$ an adjustable delay between the L and Z fields, and $A^{(Z)}$ the Z-field amplitude, assumed constant. $U^{(ZZ)}$ is invariant in space and therefore exerts no force due to its vanishing gradient. Due to the small photoacoustic conversion efficiency, $U^{(LL)}$ is negligible. Finally, the mixed potential reads $U^{(ZL)} = A^{(Z)} \langle \cos(k_z z + \omega\tau - \omega t) E^{(L)} \rangle$, with $E^{(L)} = f_1 \frac{p^{(L)}}{\rho_0 c_0^2} - \frac{3}{2} f_2 \frac{v^{(L)}}{c_0}$. After some trigonometry, we get:

$$U^{(ZL)} = A^{(Z)} \left[ \cos(\omega\tau) A_c^{(L)} - \sin(\omega\tau) A_s^{(L)} \right] \qquad 4$$

With the cosine and sine-components of the spatial-information dimensionless L-field $A_c^{(L)}(x,y,z) = \langle \cos(k_z z - \omega t) E^{(L)} \rangle$ and $A_s^{(L)}(x,y,z) = \langle \sin(k_z z - \omega t) E^{(L)} \rangle$. The other two variables $\omega\tau$ and $A^{(Z)}$ are independent of space. We note that the potential variation with $\omega\tau$ is not uniform in space, rather it is a projection between in-phase and out-of-phase components of the L-field relatively to the Z-field.

We obtain the force component in direction x by deriving Eq. 2: $F_x = A^{(Z)} \left[ \cos(\omega\tau) \partial_x A_c^{(L)} - \sin(\omega\tau) \partial_x A_s^{(L)} \right]$. Using similar trigonometry methods, we get:

$$F_x = A^{(Z)} \cos(\omega\tau - \phi_x) f_x \qquad 5$$

With $f_x = \sqrt{\left(\partial_x A_c^{(L)}\right)^2 + \left(\partial_x A_s^{(L)}\right)^2}$, $\cos\phi_x = \frac{\partial_x A_c^{(L)}}{f_x}$ and $\sin\phi_x = \frac{\partial_x A_s^{(L)}}{f_x}$.

The z-component of the force is omitted because the particles would either be pushed to the top of the channel or the bottom

## 2. Experimental setup
### 2.1 Fabrication of the laser-guided acoustic tweezers

The laser-guided acoustic tweezers are a hybrid transducer consisting of a reusable, transparent electroacoustic transducer unit (ETU) and a spectrally selective photoacoustic transducer unit (PTU).

The fabrication of the ETU starts with a 500 µm thick Y- 36° cut of LiNbO$_3$ crystal polished on both sides. The substrate thickness sets a resonance frequency of 7.6 MHz according to the longitudinal wave velocity. First, the substrate is coated with 180 nm thick indium tin oxide (ITO) on both side via Physical Vapor Deposition (PVD). The sputtered ITO behaves as transparent top and bottom electrodes. Next, the substrate is diced into 25 mm×25 mm dimensions using a laser cutting machine.

The PTU is prepared by coating a gold nanoparticle polydimethylsiloxane (PDMS) on top of the ETU. It begins by coating a 75 µm thick PDMS layer on the ETU. This is done by spin-coating PDMS (base to curing agent weight ratio of 10:1) on the surface of the ETU (1150 rpm, 30 s). The ETU is then degassed for 30 min and cured at 60℃ in an oven overnight.

Next, the generation layer of the PTU is obtained by synthesizing in-situ gold nanoparticles on the PDMS layer. The ETU coated with PDMS is placed in a crystallizing dish and immersed into 0.015 mol L$^{-1}$ tetrachloroauric acid (TCA) isopropanol solution. The crystallizing dish is sealed with parafilm to prevent evaporation, covered with aluminum foil to block light, and left to incubate for 48 h at room temperature. It is then rinsed with water, and dried with nitrogen. This completes the fabrication of the laser-guided acoustic tweezers.

### 2.2 Fabrication of the manipulation chamber

To avoid air currents and other disturbances during experiments, the ultrasonic particle manipulation is done in a 1 mm wide and 200 μm deep square microfluidic manipulation chamber. The chamber is made of PDMS and is a fabricated by soft-lithography by replicating an SU8 mold.

For the mold fabrication, a 200 μm thick SU-8 (SU-8 2075) film is spin coated on a clean 4-inch silicon wafer (500 rpm for 10 s, and then 1300 rpm for 30 s). Upon soft-baking at 65℃ for 7 min and 95 ℃ for 40 min, SU-8 is exposed to UV light for 24 s (with exposure energy of 320 mJ/cm2) followed by a post-bake at 65℃ for 5 min and 95 ℃ for 15 min. The post-baked mold is developed in photoresist developer for 17 min to obtain the designed mold shape.

To prepare the channel, PDMS (base to curing agent weight ratio of 10:1) is poured onto the SU-8 mold, and degassed for 30 min. The mixture is baked in an oven at 70 ℃ for 2 h. Then, the cured PDMS is peeled off from the mold and punched. The chamber is then bonded with the laser-guided tweezers by lightly pressing it against the PTU (PDMS is naturally sticky). After use, the PDMS channel is discarded to prevent contamination, and the laser-guided tweezers are immerged in ethanol to wash for reuse.

### 2.3 Synthesis of arbitrary microscale optical patterns and photoacoustic field resolution

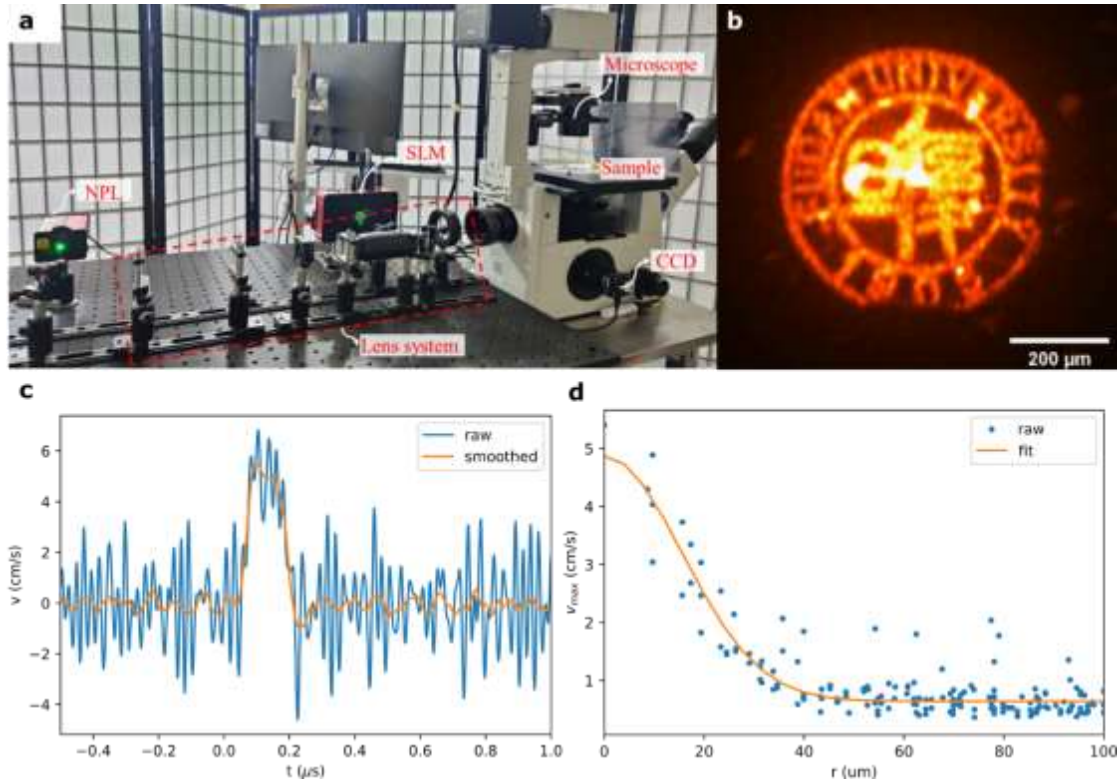

**Figure 5:** *(a)* **laser excitation system.** *(b)* *microscale optical image obtained by microscopic holography.* *The 520 nm laser beam is spatially modulated by the SLM and sent into the microscope objective. The light is observed using 200 nm fluorescent particles. (c,d) displacement speed during photoacoustic excitation of a single focused spot using a 10×. Pulse duration was set 129 ns. (c) time-dependent vertical displacement speed at the center of the laser spot. (d) spatial distribution of displacement speed depending on the distance from the center of the spot, and fit with a Gaussian bell curve. The fitted spot size is 22 µm.*

Microscale arbitrary light patterns are synthesized in a similar way to holographic optical tweezers [5]. Briefly, the light of a laser beam is circularized and expanded by a series of cylindrical lenses and beam expander, and is then directed to a spatial light modulator (SLM), where the wavefront phase is adjusted to form some desired image at an infinite distance from the SLM. The light is then guided into the illumination backport of an inverted microscope. Using a lens, the beam is conjugated to microscope backport, so that the image is focused at infinity in the absence of microscope objective. When using a microscope objective, the light pattern appears at the focal plane of the microscope. The average laser power reaching the sample was measured to be 1.25 mW (400 mW peak power).

*Figure 6.*b illustrates the formation of arbitrary light patterns by showing 200 nm diameter fluorescent particles in solution illuminated by a light pattern forming Fudan University Logo.

Focusing a 129 ns laser pulse on a single spot through a 10× objective, we get a nearly constant speed of 4.9 cm/s at the center of the laser spot. The vibration amplitude decays by 1/e within a 22 µm distance, which is defined as the spot size.

A rough estimate of the energy efficiency of the photoacoustic conversion is then obtained as $\eta = \frac{E_L}{E_{laser}}$, where $E_L$ is an order of magnitude estimate of the areal energy density of the L-wave, and $E_{laser}$ is an estimate of the areal energy density of the laser pulse. We have measured that 27% of the light from the laser reaches the sample. Based on the manufacturer data of 186 nJ/pulse and the photoacoustic spot size of 22 µm, we estimate the fluence to 3.3 mJ/cm² ($2.5\times 10^4$ W/cm²). The acoustic power is given by $\Pi = pv$. While the relation is generally not trivial, we consider here a plane wave to get an order of magnitude estimate. Accordingly, $\Pi = Zv^2$, with $Z = \rho_0 c_0$ the acoustic impedance of the medium where the wave propagates. Since the wave only exists for half an acoustic period **Figure 6.**c), we get an acoustic energy density of 16 nJ/cm². Therefore, the efficiency $\eta = 10^{-7}$.

### 2.4 Experimental acoustic fields

The Z-wave is generated by the thickness mode of the LiNbO$_3$ transducer. We checked the spatial uniformity of the beam amplitude by measuring the vibration amplitude of the substrate at three different locations several millimeters apart (**Figure 6**). Because the 50 W amplifier (50W1000B) used during manipulation was not transportable to the characterization facilities, these experiments have been carried with another less powerful amplifier (LZY-22+, minicircuits, gain 43 dB) that yield 18 W when using 300 mV$_p$ input. Reverting to 50 W amplifier (gain 47 dB) with the 316 mV$_p$ input used in our experiments, we estimate the displacement to be 1.7 nm.

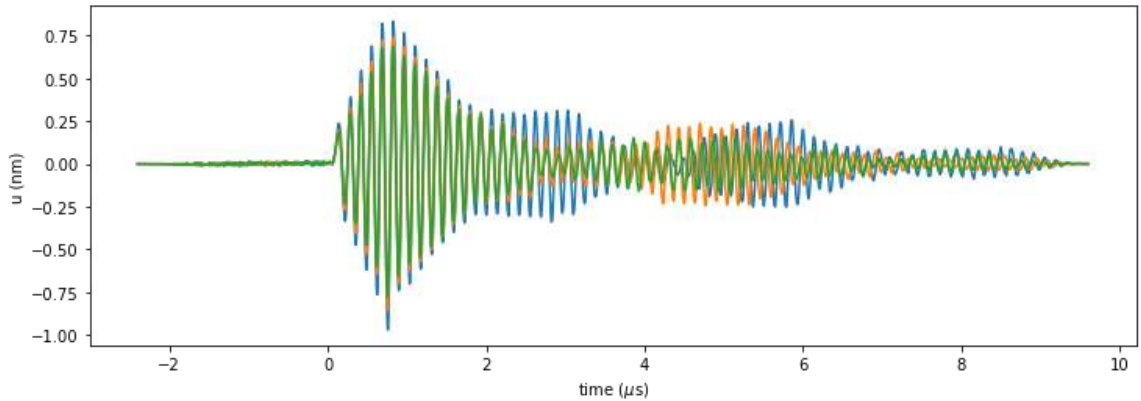

**Figure 6: Uniformity of the electroacoustic (Z) field.** Vertical mechanical displacement measured by laser vibrometry on the LiNbO$_3$ piezoelectrical transducer at three different random locations spaced apart by several millimeters. A reduced transducer power (18 W) was used for vibrometry measurements. During the first 5 periods (0 - 0.7 µs), the thickness mode resonance builds up, then decays. The coda that appears after 2 µs is likely due to reflections of transverse waves at the edge of the substrate, which should be damped by the loading of water during manipulation experiments. The photoacoustic excitation is triggered at t ≈ 0.7 µs when the vibration field is homogeneous.

### 2.5 Synchronization of electroacoustic and photoacoustic waves

The laser-guided tweezers are made of an electroacoustic transducer unit (ETU) and a photoacoustic transducer unit (PTU) that need to be powered and synchronized. Therefore, the experimental setup combines a laser excitation system (LES, to power to the PTU), an electric excitation system (EES, to power the ETU), and a trigger system (TS) to control the delay between both of them.

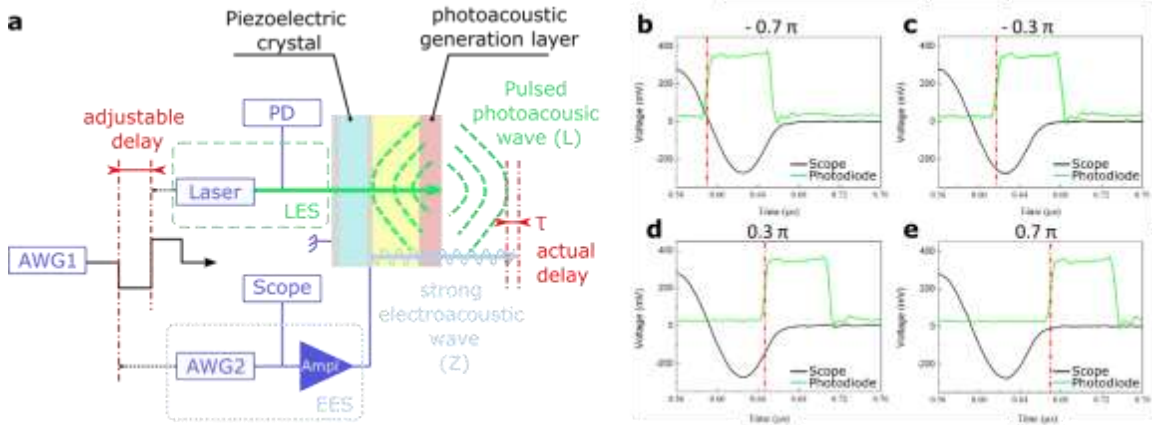

**Figure 7: Trigger system.** (a) The photoacoustic (L) and electroacoustic (Z) waves are generated by a laser excitation system (LES) and an electric excitation system (EES). These two systems are synchronized by an arbitrary waveform generator (AWG1). The EES is triggered by the falling edge of a square wave from AWG1, and the LES by the rising edge. Adjusting the duty ratio enables choosing the delay $\tau$ between the waves. (b-e) Laser intensity signal measured by a photodiode (PD) and electrical excitation signal from AWG2 measured by an oscilloscope (Scope) for various values of $\omega\tau$. We note that the piezoelectric is a resonant element and therefore the piezoelectric vibration magnitude is the highest immediately after excitation stops (see **Figure 6**).

The LES is has been described in Methods 2.3.

The EES is a Picoscope 5000 used as arbitrary function generator. It sends a sinusoidal train (five periods, voltage is 316 mV, frequency is 7.6 MHz) to a 50 W power amplifier (AR worldwide 50W1000B).

The EES and LES are synchronized using the square wave from another arbitrary function generator (Instek MFG-2260M). The falling and rising edge trigger the EES and LES, respectively. The period of the square waves matches the maximum pulse repetition frequency of the laser and the duty ratio is adjusted to tune the delay $\tau$ between the two waves.

### 2.6 Manipulation medium

The manipulation medium is a buoyancy-neutral saline solution composed of 7.8% (w/w) NaCl. To prevent the particles from sticking to the PDMS surface, 0.5% (w/w) Triton X-100 surfactant, 1% (w/w) polyethylene glycol (PEG 400) are added. Finally, in order to visualize the laser spot, 200 nm fluorescent polystyrene particle (excitation wavelength of 532 nm, and emission wavelength of 610 nm) are dispersed in the manipulation medium.

The particles to be manipulated are dispersed at very low concentration in the solution. 40 μL of the particle suspension is then injected into the channel using a micropipette. After some time, the pressure between the channel inlet and outlet are balanced and the flow stops. The acoustic manipulation then begins.

### 3. Measurement of the trapping force

The images or videos in our experiments are captured by a camera (CCD) and analyzed using imageJ.

The SLM is programmed to periodically switch between two diffraction gratings to make the beam move up and down every 10 s. We then record videos of the particle movements (supplementary video 1, 2, 3, and 4) for various phase differences ωτ.

The force is assumed to act only radially, and is deduced from the displacement by assuming Stokes drag (which underestimates the drag force and therefore provides a lower bound for the trapping force). The trapping force is computed as:

$$F = \alpha \frac{L}{\Delta t} \qquad 6$$

with L the distance travelled by the particle and $\Delta t$ = 10 s the time the particle is allowed to travel. $\alpha = 3\pi\mu d_p$ is the Stokes drag coefficient, with $\mu$ the fluid viscosity and $d_p$ the particle diameter.

As mentioned earlier, the Stokes drag is one of several dissipative forces acting on the particle. Therefore, Eq. 2 is a lower bound for the manipulation force. Other dissipative effects include wall effects (such as the bottom of the channel), that can amplify the drag force up to 8 times [32]; other hydrodynamic effects (added mass, Basset force) affecting the dynamics of the particle; and surface forces such as adhesion may also slow down the particles.

### 4. Measurement of the selectivity

The selectivity is measured by tracking several particles simultaneously as the laser is alternately moved from left to right (subscript R) and backward (subscript L). The force is assumed to be axisymmetric and its profile is assumed be the same as the one returned by the finite element model, but its amplitude is unknown, and is therefore fitted to a coefficient $\beta$. Our model also accounts for some small residual flow $v_F$ in the channel, with an unknown direction and amplitude. Yet, this flow is assumed to be independent of the laser position.

Assuming that for a given cycle j, the laser spot in on the right ($\mathbf{r_R}$ denotes the laser position, $\mathbf{r_L}$ would be used if the laser were on the left), then particle i located at $r_i$ obeys the following motion in the x-direction:

$$v_{F,x}(\mathbf{r_j}) + \frac{\beta}{\alpha}\mathbf{F}\left(||\mathbf{r_R} - \mathbf{r_j}||\right)\cos\theta_{R,j} = v_{p,x}^{i,j} \qquad 7$$

and in the y-direction:

$$v_{F,y}(\mathbf{r_j}) + \frac{\beta}{\alpha}\mathbf{F}\left(\left\|\mathbf{r_R} - \mathbf{r_j}\right\|\right)\sin\theta_{R,j} = v_{p,y}^{i,j} \qquad 8$$

Where $\theta_{R,j} = \arctan 2\left((\mathbf{r_R} - \mathbf{r_j})\cdot\mathbf{e_y}, (\mathbf{r_R} - \mathbf{r_j})\cdot\mathbf{e_x}\right)$ is the relative angle between the particle location and the laser, $v_{p,x}^{i,j}$ is the particle (i) displacement speed during cycle j, and $v_{F,x}$, $v_{F,y}$ and $\beta$ are the unknowns, which represent the x and y components of the drift flow for each particle (assumed constant over the left-right cycles of the laser), and the amplitude coefficient for the force (shared for all particles), respectively. 126 experimental datapoints (corresponding to three particles) are fitted to find these 7 unknowns.

To show that our fitting approach places strong constraints on the acoustic field, the result obtained when fitting other common radiation force patterns are shown below. The theoretical profile from the finite element model fits best with experimental data.

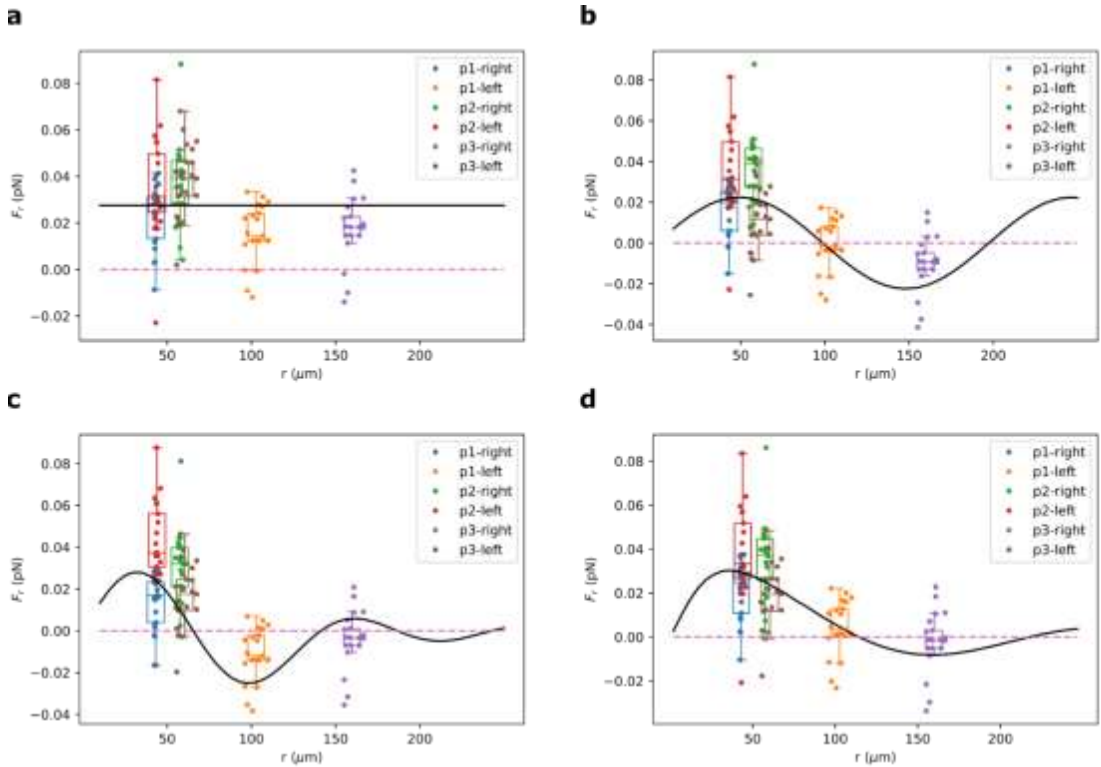

**Figure 8: fit of the experimental data against different force profiles.** (a) constant radial force, (b) sinusoidal force, (c) acoustic vortex tweezers, (d) theoretical acoustic radiation force. The boxes indicate the lowest and highest quartiles (they contain 50% of the experimental datapoints), the whiskers indicate the limit for outsiders (1.5 times the interquartile distance). All the alternative models miss at least three boxes (containing 50% of the measured points). The simulated model misses only the last box by a small margin.

## 5. Finite element simulation of the transient acoustic radiation force

The acoustic radiation force was computed from Eq. 2 and 3 based on the acoustic pressure simulated with COMSOL Multiphysics "pressure acoustics transient" module:

$$\frac{1}{c_0{}^2}\frac{\partial^2 p}{\partial t^2} = \Delta p. \qquad 9$$

The velocity was deduced from the pressure as $v = -\int \frac{1}{\rho}\nabla p\, dt$, and the time-average operator was used in post-processing to obtained the time-averaged acoustic radiation potential.

# Acknowledgements

This work was supported by the National Natural Science Foundation of China with Grant Nos. 12004078, 51950410582, and 61874033, and the State Key Lab of ASIC and System, Fudan University with Grant Nos. 2021MS001 and 2021MS002.

# Authors information

### Affiliations

> **State Key Laboratory of ASIC and System, School of Microelectronics, Fudan University, Shanghai 200433, China**
>
> Qing Wang, Shuhan Chen, Jia Zhou and Antoine Riaud

### Contributions

A.R. initiated the project. A.R and Q.W. designed the experiment. Q.W. produced the transducers and conducted the experiment. A.R., Q.W. and S.C. analyzed the data. A.R. and S.C. conducted the simulations. A.R. and J.Z. directed the research project. All the authors wrote the manuscript together.

### Corresponding author

Correspondence to [Jia Zhou](#) and [Antoine Riaud](#).

## Authors information

**Competing interests**

The authors filed two patent applications related to laser-guided acoustic tweezers.

All photographs were taken by the authors.